\documentclass[10pt,letterpaper]{article}
\usepackage{opex3}

\newcommand{\br}{{\bf r}}
\newcommand{\bE}{{\bf E}}

\newcommand{\bG}{{\bf G}}
\newcommand{\bp}{{\bf p}}

\newcommand{\bu}{{\bf u}}

\newcommand{\dd}{{\mathrm{d}}}

\newcommand{\dddd}{{\mathrm{d}}^3}

\newcommand{\Imag}{\mathrm{Im} \, }

\newcommand{\Tr}{\mathrm{Tr} \, }
\newcommand{\nm}{\,\mathrm{nm}}

\begin{document}

%%%%%%%%%%%%%%%%%% title page information %%%%%%%%%%%%%%%%%%
\title{Towards a full characterization of a plasmonic nanostructure with a fluorescent near-field probe}

\author{V. Krachmalnicoff$^1$, D. Cao$^1$, A. Caz\'e$^1$, E. Castani\'e$^1$, R. Pierrat$^1$, N. Bardou$^2$, S. Collin$^2$, R. Carminati$^1$, and Y. De Wilde$^1$}

\address{${}^1$Institut Langevin, ESPCI ParisTech \& CNRS UMR 7587, 1 rue Jussieu, 75005 Paris, France\ ${}^2$Laboratoire de Photonique et Nanostructures (LPN-CNRS), Route de Nozay, 91460 Marcoussis, France}

\email{valentina.krachmalnicoff@espci.fr}
\email{yannick.dewilde@espci.fr} %% email address is required

% \homepage{http:...} %% author's URL, if desired

%%%%%%%%%%%%%%%%%%% abstract and OCIS codes %%%%%%%%%%%%%%%%
%% [use \begin{abstract*}...\end{abstract*} if exempt from copyright]

\begin{abstract}
We report on the experimental and theoretical study of the spatial fluctuations of the local density of states (EM-LDOS) and of the  fluorescence intensity in the near-field of a gold nanoantenna. EM-LDOS, fluorescence intensity and topography maps are acquired simultaneously by scanning a fluorescent nanosource grafted on the tip of an atomic force microscope at the surface of the sample. The results are in good quantitative agreement with numerical simulations. This work paves the way for a full near-field characterization of an optical nanoantenna.
\end{abstract}


%%%%%%%%%%%%%%%%%%%%%%% References %%%%%%%%%%%%%%%%%%%%%%%%%

%%%%%%%%%%%%%%%%%%%%%%%%%%  body  %%%%%%%%%%%%%%%%%%%%%%%%%%
\section{Introduction}
Much interest has recently been paid to plasmonic nanostructures due to the capacity they offer to enhance light-matter interaction of elementary dipoles such as fluorescent molecules and quantum dots. This encompasses three different mechanisms which are often hard to disentangle. (1) Light absorption can be enhanced by an optical antenna, leading to an increased effective absorption cross section. This can be advantageously used in photodetection and photovoltaics \cite{Novotny_NatPhot2011}. (2) Plasmonic nanostructures may also induce significant changes in the spontaneous emission dynamics by the Purcell effect. As an example, large Purcell factors associated to electromagnetic modes strongly localized in subwavelength volumes have been reported at the surface of disordered thin metallic films, which opens new perspectives such as the investigation of strong coupling in a regime where high Purcell factors coexist with high absorption \cite{Krachmalnicoff_PRL2010, Castanie2012, Canneson_PRB2011}. (3) A change of fluorescence intensity is also expected when a dipole is in near-field interaction with a plasmonic nanostructure. In an optical antenna, a field enhancement occurs, but it is often obscured by non-radiative processes leading to fluorescence quenching.

The local density of electromagnetic states (EM-LDOS) is the basic quantity which governs these three mechanisms. As it rules the energy stored in all available modes, it provides a direct measurement of the probability of spontaneous light emission. The decay rate of a fluorescent dipole is proportional to the EM-LDOS. While in vacuum the EM-LDOS is homogeneous, it is known that it can be significantly affected by the proximity of an interface or a nanostructure \cite{Joulain_PRB2003, NovotnyBook}. Nanostructures made of dielectric materials such as microcavities have been shown to induce an increase or a decrease of the fluorescence decay rate \cite{Vahala_Nature2003}. Several groups have also reported spatial variations of the EM-LDOS in disordered photonic media, based on statistical measurements of the fluorescent emission rate from a large collection of dipole emitters \cite{Krachmalnicoff_PRL2010,Birowosuto_PRL2010,Sapienza_PRL2011}. In such systems, the direct observation of modes strongly localized in subwavelength volumes is crucial as it can provide an important signature of Anderson localization and opens the route to cavity quantum electrodynamics in systems which are inherently disordered \cite{Sapienza_Science2010}. Optical nanoantennas constitute another example of systems in which electromagnetic field can be controlled on a nanometer scale, producing an environment which can modify the amount of energy emitted by molecules and their direction of emission \cite{Novotny_NatPhot2011,Curto_Quidant_Science2010,Greffet_Science2005,Greffet_PRL2010,Agio_NanoScale2012, Vandembem2009, Vandembem2010}.

A key issue to probe the EM-LDOS experimentally is that the detection process should have the same response for all modes. Several methods have been proposed to map the spatial variations of the EM-LDOS on photonic nanostructures, among which measuring the thermal emission in the near-field \cite{DeWilde_Nature2006}, measuring the ``forbidden light" signals from the aperture of a near-field scanning optical microscope \cite{Dereux}, or using a scanning electron beam as a point dipole source \cite{Sapienza_NatMat2012}. The EM-LDOS can be directly inferred from measurements of the spontaneous fluorescence decay rate of a single nanoemitter in its local environment, $\Gamma=1/\tau$, where $\tau$ is the fluorescence lifetime of the emitter \cite{Krachmalnicoff_PRL2010,NovotnyBook}. An emitter constituted by several randomly oriented dipoles probes simultaneously the EM-LDOS in the three spatial directions. One dimensional maps of the decay rate were initially obtained by scanning a gold bead near single isolated fluorescent molecules fixed on a substrate \cite{Kuhn_PRL2006,Anger_PRL2006}. Recently, a decrease of the fluorescence lifetime was measured in the reverse situation, when scanning a fluorescent bead across a 250 nm diameter silver rod, pointing to an increased EM-LDOS due to the existence of plasmonic modes on the rod \cite{Frimmer_PRL2011}.
   
While a detailed knowledge of the EM-LDOS is clearly required, it is not enough to provide a full characterization of a system involving dipoles coupled with plasmonic nanostructures. Local changes of fluorescence intensity depend on other parameters such as the radiative and non-radiative part of the EM-LDOS and the local field enhancement factor \cite{Wenger2008}. To characterize a plasmonic antenna, one needs at least to measure both the EM-LDOS and the fluorescence enhancement factor at the nanometer scale in the near-field of the antenna. 

In this paper, we present the use of a fluorescent 100-nanometers bead grafted at the apex of a scanning probe for the near-field investigation of an optical antenna. Two dimensional (2D) maps of the decay rate and intensity of the fluorescent bead are obtained with subwavelength resolution. The experimental near-field images of fluorescence intensity and EM-LDOS provide complementary information that are both needed to characterize the optical response of a nanoantenna. A good understanding of these maps is achieved by comparison with  numerical simulations,  which also allows one to analyze the optical response of the fluorescent near-field probe used to perform the experiments.

The decay rate averaged over the orientations of the transition dipole $\bp$ reads
\begin{equation}
\Gamma = \frac{\pi \omega}{3 \epsilon_0 \hbar} |\bp|^2 \, \rho(\br,\omega),
\end{equation}\label{eq:Gamma}
where $\rho(\br,\omega)$ is the EM-LDOS. Hence, measuring the fluorescence lifetime $\tau=1/\Gamma$ is a direct way to probe the EM-LDOS.

If the molecules are far from saturation, the fluorescence signal reads
\begin{equation}
\label{fluorescence_signal}
S = C  \left [ \int_{\Omega} \eta(\bu,\omega_{fluo}) \dd \Omega \right ] \, \sigma(\omega_{exc}) \, K^2(\omega_{exc}) \, I_{inc}.
\end{equation}
In this equation, $\eta(\bu,\omega_{fluo})$ is the apparent quantum yield for a detection in direction $\bu$, and $\Omega$ is the solid angle
of the detection objective. The apparent quantum yield is defined as
$\eta(\bu,\omega_{fluo}) = \Gamma_R(\bu)/\Gamma$ , with $\Gamma_R(\bu)$ the directional radiative decay rate and $\Gamma=1/\tau$ the full decay rate.
$\omega_{exc}$ and $\omega_{fluo}$ are the absorption and emission frequencies of the molecules. 
The constant $C$ is a calibration parameter of the detection (that accounts for transmissivity of filters, detector efficiency, etc), $\sigma(\omega_{exc})$ is
the absorption cross-section of the bare fluorescent beads, $I_{inc}$ is the incident laser intensity and $K^2(\omega_{exc})$ is the local-intensity 
enhancement factor. In terms of the fluorescence signal, the product $F=\left [ \int_{\Omega} \eta(\bu,\omega_{fluo}) \dd \Omega \right ] \,  K^2(\omega_{exc})$ is the fluorescence enhancement factor and drives the contrast of the images. 

\section{Experimental results}
\begin{figure}[h!]	
	\begin{center}
		\includegraphics[width=7cm]{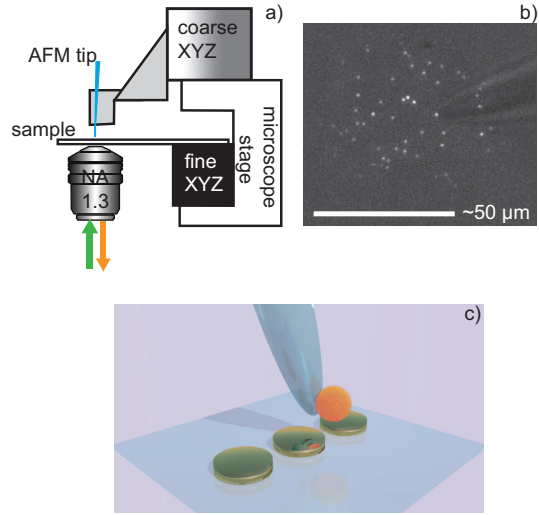}
		\caption{\label{fig:experimental_setup} a) Sketch of the experimental setup. The active AFM tip is mounted on a piezoelectric system allowing the positioning of the tip within the laser diffraction limited spot. The excitation and fluorescence photons are respectively focused and collected from the same high NA objective. The sample can be moved on the XY plane to perform the fluorescence intensity, EM-LDOS and topography maps.  b) Wide-field fluorescence image of the beads spread on the microscope coverslip. The tip can be seen on the right side of the image. One bead is grafted on the apex of the tip (see text and movie in supplemental data). c) Artist view of the active AFM tip scanning the near-field of the gold nanoantenna.}
	\end{center}
\end{figure}

Active fluorescent probes which allow near-field intensity measurements have been reported \cite{Lewis1990,Michaelis2000,Kuhn2001,Aigouy2003,Huant2010,Wrachtrup2008,Rondin2012}, but a scanning probe based on a fluorescent nanobead in which {\it both} the fluorescence decay rate and the fluorescence intensity are simultaneously measured with subwavelength resolution is still very uncommon \cite{Frimmer_PRL2011}. Such a probe constitutes a major advance in the field of nano-optics as it provides a direct access to a map of the EM-LDOS and the intensity. Its practical realization firstly requires manipulating the bead to graft it at the apex of the tip of the scanning probe, and then recording its fluorescence lifetime and intensity at a rate compatible with the acquisition of 2D images. The experimental setup which we have designed to this aim is sketched in Fig.\ref{fig:experimental_setup}a. It consists of a homebuilt atomic force microscope (AFM) at the tip of which a fluorescent nano-bead is fixed. The AFM is mounted on the stage of an inverted confocal  fluorescence microscope equipped with a white pulsed laser (Fianum SC450), an avalanche photodiode (Micro Photon Devices), and a time correlated single photon counting (TCSPC) electronics for fluorescence lifetime measurements (Picoquant HydraHarp 400) \cite{Krachmalnicoff_PRL2010}. The scans are performed on the sample stage using a three axes (xyz) piezoelectric assembly (Piezosystem Jena). The tip is a pulled silica fiber with an apex of $\approx 100$~nm diameter, mounted at the extremity of one arm of a quartz tuning fork excited by a dither piezoceramic. Stepper motors (Smaract GmbH) allow three dimensional displacements of the tip to perform the coarse approach and the fine positioning of the tip within the tightly focused laser spot produced by a high numerical aperture (NA=1.4) microscope objective. The tip oscillates laterally in shear mode at a height of approximately $10$~nm above the surface of the sample thanks to a feedback electronics \cite{Karrai_PRB2000}. The optical probe consists of a $100$~nm diameter polymer bead filled with fluorescent dye molecules (Invitrogen Red Fluospheres) which is grafted at the extremity of the silica tip. Fluorescent beads with a diameter of $100$~nm are initially spin-coated on a glass coverslip with a separation of  $5$ to $10$~$\mu \mathrm{m}$ between them. Wide-field illumination of the coverslip is initially used to excite simultaneously the fluorescence of several beads (see Fig.\ref{fig:experimental_setup}b), which are imaged with an EM-CCD camera (Andor iXon 897) mounted on an extra port of the inverted microscope. The silica tip is brought at the glass surface. The grafting of a single fluorescent bead on its apex is achieved by moving the bead in contact with the tip while monitoring the process in real time with the CCD camera (see movie in supplemental data). Fig.\ref{fig:experimental_setup}b shows an optical fluorescence image of the beads spread on the glass coverslip, one of them being grafted on the silica tip visible on the right side of the image. Remarkably, there is no glue involved in the grafting which is successful in about $50$\% of the attempts and can last up to several days. Once the grafting procedure is achieved, the nano-bead is brought close to the structure under study, the laser ($\lambda_{exc} =560$~nm) is tightly focused on the bead in order to excite its fluorescence and the sample is scanned. Note that, as the grafting is achieved by lateral displacements of the bead towards the tip (see supplemental data), the bead is expected to be laterally displaced with respect to the tip apex, as schematically sketched in Fig.\ref{fig:experimental_setup}c.

As a first example of the investigation of a nanoantenna, we have studied a linear chain of three $150$~nm-diameter gold nano-disks separated by $50$~nm gaps on a glass substrate \cite{Suck_OL2011}. According to numerical simulations discussed later in the paper, such a nanostructure is expected to produce significant variations of the EM-LDOS and of the fluorescence enhancement factor on a scale of several tens of nanometers, which we have observed with the fluorescence scanning probe. The disk chains are fabricated by electron beam lithography on a glass microscope coverslip. Each disk is made of a $2$~nm-thick wetting layer of chromium and of a $30$~nm-thick layer of gold. 

\begin{figure}[h!]	
	\begin{center}
		\includegraphics[width=7cm]{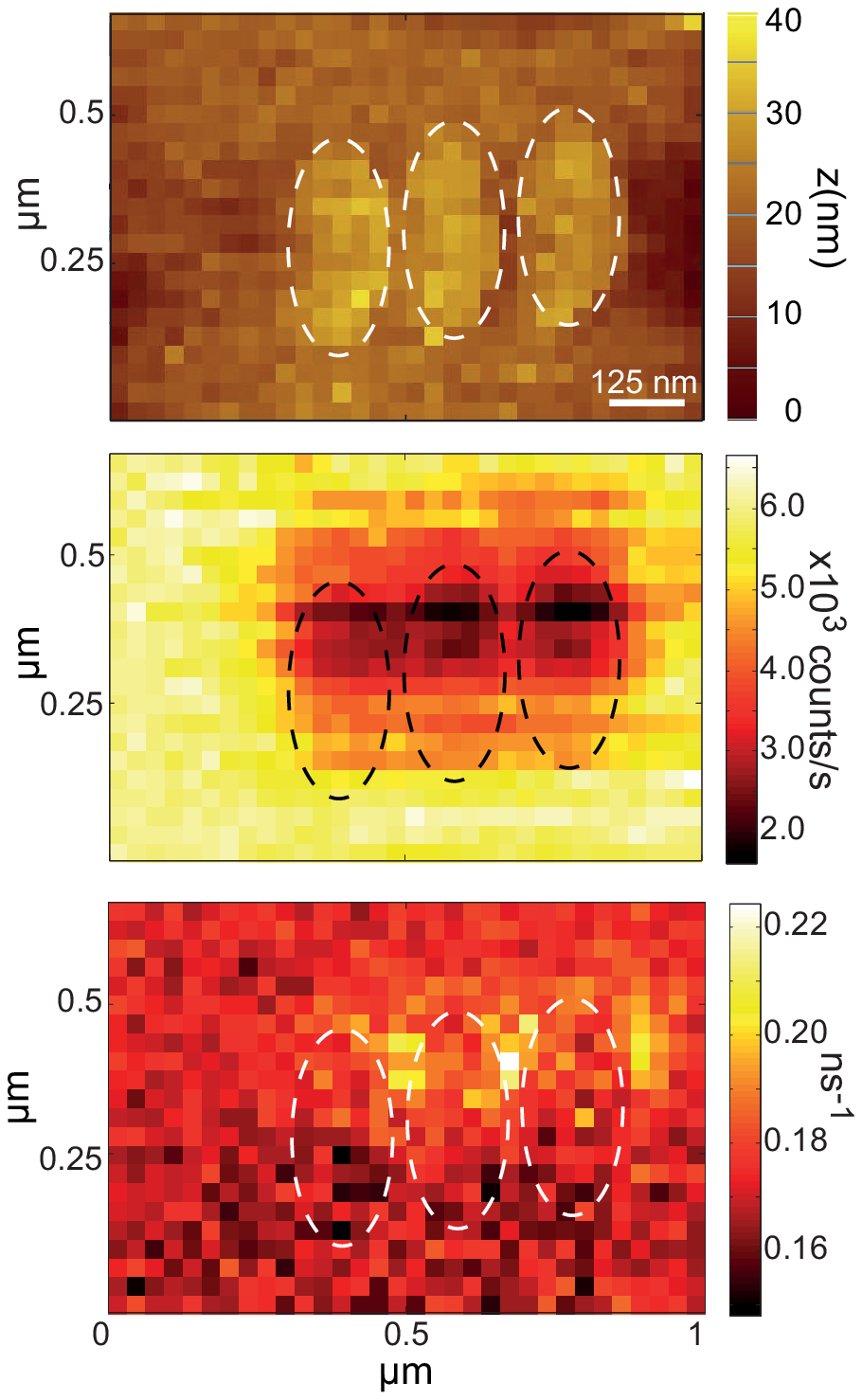}
		\caption{\label{fig:experimental_maps} Experimental results: a) Topography of the sample. b) Fluorescence intensity map. c) Decay rate (EM-LDOS) map. The contour of the topographic relief (dashed line), as measured by the active AFM asymmetric probe (see text), is reported on the three maps to guide the eye.}
	\end{center}
\end{figure}

Three signals are simultaneously recorded during the scans as a function of the position of the fluorescent probe at the surface of the sample, producing three different images of the nanoantenna: (1) the topography (AFM image, see Fig.\ref{fig:experimental_maps}a); (2) the fluorescence intensity, which corresponds to the integral of the decay rate histogram (Fig.\ref{fig:experimental_maps}b); and (3) the spontaneous decay rate (Fig.\ref{fig:experimental_maps}c), obtained after measuring, at each point of the scan during $1$~s, the histogram of the arrival time of the fluorescence photons with respect to the excitation laser pulse \cite{Krachmalnicoff_PRL2010}. This long acquisition time is a major difficulty to record 2D maps of the decay rate on a very small area with a scanning probe, because the system has to be stable with nanometric precision against thermal and mechanical drifts. This explains the low number of pixels ($32\times22$) of the experimental images.
% Fig 1 : Experimental setup

% Fig 2 : Experimental results
The topography of the sample shows three distinct objects with a height of $30$~nm each. The measured structures seem not to be circularly shaped as expected, but rather elliptical. This is because the measured topography is given by the convolution of the topography of the real object (three disks) with the shape of the probe used to make the scan, here the AFM tip on which a fluorescent bead is grafted sideways. Due to the lateral shift of the bead with respect to the tip apex (Fig.\ref{fig:experimental_setup}c), the topography probe presents an anisotropic shape. The size of the short axis in the AFM image of a disk is ruled by the larger object which constitutes the probe, while the size of the longer axis is related to the sum of the sizes of the bead and the tip. Hence, the topographic image confirms the lateral displacement of the bead with respect to the tip apex and allows one to affirm that the size of the tip is of the same order of the size of the bead.

The geometry of the scanning probe as sketched in Fig.\ref{fig:experimental_setup}c, is confirmed by looking at the fluorescence intensity map shown in Fig.\ref{fig:experimental_maps}b. The contour of the measured topography (Fig.\ref{fig:experimental_maps}a) is reported on this map to guide the eye (dashed lines). The intensity signal decreases in three regions, circularly shaped, situated on the upper half of the elliptical contour. Since the fluorescence signal only comes from the $100$~nm diameter dye doped bead, this confirms that the three golden disks are scanned twice, once by the bead and then by the tip. The decrease of the fluorescence intensity in correspondence of the gold structure is confirmed by the numerical simulations shown in Fig.\ref{fig:numerical_maps}. Remarkably, the numerical and experimental intensity maps present an almost quantitative agreement, showing both a contrast of about a factor $3$ of the  fluorescence intensity. Note also that the combination of the topography and the fluorescence intensity signals allows us to intuit quite precisely what is the position of the bead with respect to the tip apex.

Beyond the fluorescence intensity map, a deeper insight of the properties of the electromagnetic field at the surface of this nanoantenna is obtained by mapping the decay rate of the fluorescent bead, which is proportional to the EM-LDOS (see Eq.\ref{eq:Gamma}).  
Fig.\ref{fig:experimental_maps}c shows a decay rate map of the scanned area. It is observed that the EM-LDOS increases by about $30\%$ in three regions presenting an extension of about $60$~nm each and separated by $100$~nm.
Two of these regions are situated between the gold disks, as confirmed by numerical simulations reported in Fig.\ref{fig:numerical_maps}c and discussed below. As in the case of the fluorescence intensity map, numerical and experimental data are in almost quantitative agreement regarding the expected change of the decay rate in the region between the disks with respect to a region far away from the nanoantenna.  Note that the numerical simulations also show the presence of two other regions with an enhanced EM-LDOS, on the external sides of the nanoantenna. Experimentally, we are only able to see one of these regions, situated on the right side of the antenna. A possible explanation for this is an asymmetry of the gold structure, caused for example by a defect of the lift-off process, that would translate in an asymmetry of the structure of the electromagnetic field on the surface of the nanoantenna. Numerical calculations with asymmetric shaped nanoantennas have been done and produce similar asymmetries in the EM-LDOS images. However, since the exact shape of the nanoantenna is not accessible at the required level of resolution, having an exact matching between theory and experiment is a very speculative task and the discussion is therefore limited here to a comparison between the experimental results with numerical simulations made on an ideal antenna formed by three regularly spaced circular disks. %\emph{add a comment about the height of the bead that could not be the same on the left and right of the scan?}

\section{Numerical simulations: comparison with the experiment}
\begin{figure}[h!]	
	\begin{center}
		\includegraphics[width=7cm]{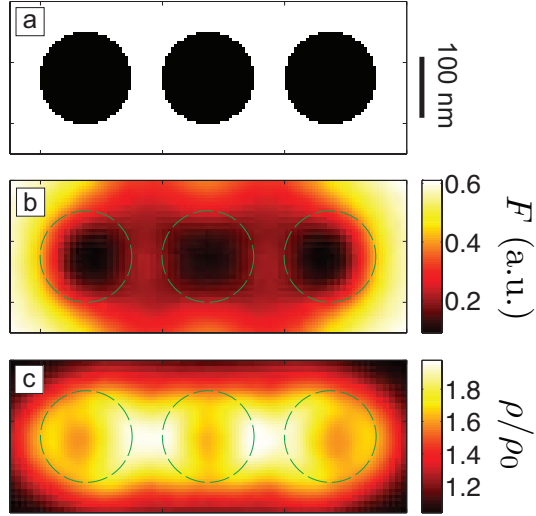}
		\caption{\label{fig:numerical_maps} (a) Top view of the topography of the discretized trimer. Note that the trimer is $30\,\mathrm{nm}$ thick, like the one used in the experiment; (b) Numerical fluorescence signal map (expressed in arbitrary units); (c) Numerical EM-LDOS map normalized to its value in vacuum.}
	\end{center}
\end{figure}

In order to analyze the experimental results precisely, we have performed exact 3D numerical simulations. We use a volume integral method based on the Lippmann-Schwinger equation
\begin{equation}
\label{lippmann_schwinger}
\bE(\br,\omega) = \bE_0(\br,\omega) + \frac{\omega^2}{c^2} \int \left[\epsilon(\br',\omega)-1\right]\bG_0(\br,\br',\omega)\bE(\br',\omega)\dddd\br',
\end{equation}

where $\bE_0$ is the incident field, $\bG_0$ the dyadic Green function of the background (vacuum in our simulations) and $\epsilon(\omega)$ is the dielectric function of gold~\cite{PalikBook}.
The numerical computation is done by a moment method, discretizing the trimer into 5 nm cells (see Fig.\ref{fig:numerical_maps}a).
 The Green function $\bG_0$ is integrated over the cell to improve convergence~\cite{Chaumet2004}. Computing the electric field under the illumination of a source dipole $\bE_0(\br , \omega) = \mu_0\omega^2\bG_0(\br,\br_0,\omega)\bp$, we can deduce the total Green function $\bG$ from $\bE(\br , \omega) = \mu_0\omega^2\bG(\br,\br_0,\omega)\bp$. 
The EM-LDOS is then given by
\begin{equation}
\label{ldos}
\frac{\rho(\br_0,\omega)}{\rho_0} = \frac{2\pi}{k_0}\, \Imag\,\Tr\,\left[ \bG(\br_0,\br_0,\omega) \right],
\end{equation}
where $\rho_0$ is the EM-LDOS in vacuum, given by $\rho_0=\omega^2/(\pi^2c^3)$, $k_0 = \omega/c$ and $\Tr$ denotes the trace of a tensor. 
The fluorescence enhancement factor, given by Eq.\ref{fluorescence_signal}, is driven by the product
\begin{equation}
F = \left [ \int_{\Omega} \eta(\bu,\omega_{fluo}) \dd \Omega \right ] \, K^2(\omega_{exc}).
\end{equation}
The apparent quantum yield of the molecules can be deduced from the total Green function $\bG$~\cite{Caze2012}.
For sake of simplicity, we do not take into account any intrinsic non-radiative decay rate that would account for internal losses.
Moreover, in our simulation, we do not integrate over the directions $\bu$ of collection but we only consider the direction orthogonal to the plane of the antenna. 
To compute $K(\omega_{exc})$ we solve Eq.\ref{lippmann_schwinger} under the illumination of a plane-wave with normal incidence. 
We average on two orthogonal polarizations to mimic the fact that the incident laser beam in the experiment is not polarized.

% Numerical results
In Fig.\ref{fig:numerical_maps}, we present the results of our calculations.
 Every point of each map is averaged over 100 positions of the emitter randomly chosen inside a $100$~nm diameter sphere, to mimic the experimental fluorescent beads. 
The distance between the top of the trimer and the bottom of the bead is fixed at $d=20$~nm. The excitation and emission wavelengths have respectively been set to $\lambda_{exc}=560\,\mathrm{nm}$ and $\lambda_{fluo}=605\,\mathrm{nm}$. 
 The dimensions of the trimer are the same as the experimental ones.

The general trends observed in the experimental maps are in excellent agreement with the simulations.
The decay rate exhibits two major hot-spots in the two gaps between the disks, and two minor ones on both sides.
The fluorescence signal is significantly reduced when the molecules are on top of a disk.
The contrast of both maps are in nearly quantitative agreement with the experimental results, which confirms us in the idea that the order of magnitude of the distance between the bead and the trimer is sound.

% Fig 3 : Numerical results

\section{Numerical study of the resolution of the EM-LDOS map}
\begin{figure}[h!]	
	\begin{center}
		\includegraphics[width=10cm]{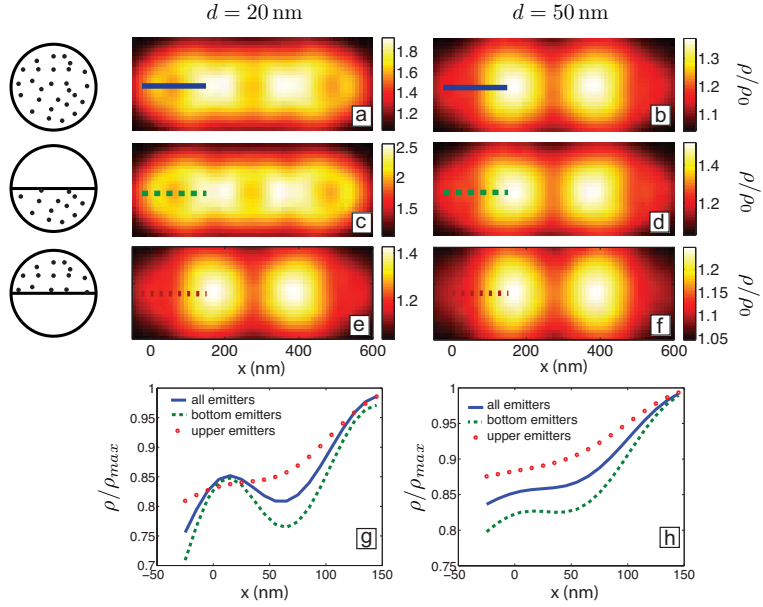}
		\caption{\label{fig:smallest_detail} Computed normalized EM-LDOS maps for two distances $d$ between the \emph{bottom} of the bead and the top of the trimer. (a-b) Average over 100 emitters randomly located in the bead; (c-d) Contribution of the 48 emitters located in the lower half of the bead; (e-f) Contribution of the 52 emitters located in the upper half of the bead. (g-h) Section view of the maps (a,c,e) and (b,d,f) respectively along the line shown on the maps. Note that in this case the EM-LDOS has been normalized by the maximum value of each map $\rho_{max}$ to quantify the contrast of the image. $\lambda_{exc} = 560\,\mathrm{nm}$; $\lambda_{fluo} = 605\,\mathrm{nm}$. Diameter of the bead: $100\,\mathrm{nm}$.}
		\end{center}
\end{figure}

One interesting feature of both the experimental and numerical EM-LDOS maps is that both seem to exhibit variations on scales well below 100$\,\mathrm{nm}$, the size of the fluorescent bead. 
To explain this phenomenon, already observed in~\cite{Frimmer_PRL2011}, we compare the contribution to the decay rate of the emitters located in the lower half and the upper half of the bead.
Fig.\ref{fig:smallest_detail}a-b show the EM-LDOS maps averaged over 100 emitters located at different positions inside a $100\nm$ bead for two distances $d$ between the bottom of the bead and the top of the trimer. Fig.\ref{fig:smallest_detail}c-d and Fig.\ref{fig:smallest_detail}e-f show the maps averaged over the emitters located respectively in the lower and the upper half of the bead. Each map is normalized by the value of the EM-LDOS in vacuum to allow for the comparison between the maps. 

A comparison between Fig.\ref{fig:smallest_detail}a and b shows that the smallest details present on the EM-LDOS map (such as for example the two EM-LDOS hot spots visible on the right and left sides of the nanoantenna) are washed out when the distance of the bead to the sample surface increases. This is a known feature in near-field optics. In fact, subwavelength details are evanescent and can only be probed by emitters in the very near-field of the system \cite{Castanie2012}. Since these details are visible on the experimental map, this study confirms that the real distance between the bottom of the probe and the sample surface is of the order of $20$~nm.

Furthermore, a detailed observation of Fig.\ref{fig:smallest_detail}a, b and c allows us to affirm that the resolution of the EM-LDOS map is not limited by the size of the bead. Indeed, the similarity between Fig.\ref{fig:smallest_detail}a and c clearly shows that the EM-LDOS map is driven by the emitters situated on the lower half of the bead. The two EM-LDOS hot spots which are visible on the right and on the left side of the nanoantenna are smeared out when considering only the contribution of the emitters populating the upper part of the sphere. 
More insight can be given by plotting the section of the EM-LDOS maps along the lines drawn in Fig.\ref{fig:smallest_detail}a, c, e. The obtained profiles are normalized by the maximum value of the corresponding map $\rho_{max}$, in order to quantify the contrast of each hotspot. They  are reported in Fig.\ref{fig:smallest_detail}g and in Fig.\ref{fig:smallest_detail}h for $d=20$~nm and $d=50$~nm respectively. For $d=20\nm$, the lateral hot-spot is clearly resolved when the EM-LDOS signal is averaged on the emitters located on the bottom of the sphere or over all the sphere, while it is washed out when the signal is averaged over the top of the sphere. Therefore the resolution of this detail is clearly due to the bottom emitters. Consequently, the effective resolution is not limited by the size of the bead but is smaller and in the case presented in this paper is of the order of $50$~nm. Interestingly, at $d=50\nm$, even if a non-monotonic behavior is observed in the profiles reported on Fig.\ref{fig:smallest_detail}h), the bottom emitters are too far away from the sample surface and the smallest details are washed out.

% Fig 4 : Smallest detail discussion

\section{Conclusion}
In conclusion, the scanning probe described in this paper has demonstrated its ability to map both the fluorescence intensity and the EM-LDOS simultaneously with nanometer accuracy. While its use was devoted here to the investigation of an optical antenna, it can be used to the investigation of the electromagnetic modes on any type of photonic nanostructure. The manipulation of a fluorescent nano-object with nanometer accuracy at the surface of a photonic nanostructure opens interesting new perspectives for studies of quantum electrodynamics, such as the investigation of the coupling of quantum emitters with plasmonic (or dielectric) devices, the characterization of electromagnetic modes on photonic nanostructures, or the search for Anderson localized modes in disordered systems.   

\section*{Acknowledgements}
We acknowledge Abdel Souilah for technical support.
This work is supported by the French National Research Agency (ANR-11-BS10-0015,"3DBROM"), by the Region Ile-de-France in the framework of DIM Nano-K and by LABEX WIFI (Laboratory of Excellence within the French Program "Investments for the Future") under reference ANR-10-IDEX-0001-02 PSL*.

\end{document}